\documentclass{pasj00}

\begin{document}
\SetRunningHead{Omukai}{Effect of pre-ionization 
in low-metallicity clouds}
\Received{}
\Accepted{}
\Published{}

\title{Do the environmental conditions affect the 
dust-induced fragmentation in low-metallicity clouds?:
Effect of pre-ionization and far-ultraviolet/cosmic-ray fields}

\author{Kazuyuki \textsc{Omukai}}
\affil{Department of Physics, Kyoto University, 
Kyoto 606-8502, Japan}
\email{omukai@tap.scphys.kyoto-u.ac.jp}


\maketitle

\begin{abstract}
We study effects of the fully ionized initial state, or pre-ionization, 
on the subsequent thermal evolution of low-metallicity 
clouds under various intensities of the external far-ultraviolet(FUV) 
and cosmic-ray(CR) fields.
The pre-ionization significantly affects 
the thermal and dynamical evolution of metal-free clouds without 
FUV/CRs by way of efficient HD formation. 
On the other hand,  
the pre-ionization effect on the thermal evolution is limited 
in very low-density regime for more 
metal-enriched clouds ([Z/H] $\gtrsim -4$) or those
under modest FUV ($\gtrsim 10^{-3}$) or CR field ($\gtrsim 0.1$ of the 
present-day Galactic disk levels). 
In any case, for $\gtrsim 10^8{\rm cm^{-3}}$, 
neither the initial ionization state nor the 
irradiating FUV strength affect the thermal evolution. 
The dust cooling is an important mechanism for 
making sub-solar mass fragments in low-metallicity gas.   
Since this fragmentation occurs at the temperature minimum by 
the dust cooling at $\gtrsim 10^{10} {\rm cm^{-3}}$, this process is 
not vulnerable either to initial ionization state or external radiation.
\end{abstract}

\section{Introduction}
The first generation of stars, or the so-called 
population III.1 stars (pop III.1 stars; O'Shea et al. 2008), are formed 
out of the primordial pristine gas with a small ionization 
degree ($\sim 10^{-4}$) left from the cosmic recombination 
(Peebles 1968; Sasaki \& Takahara 1993). 
Within minihalos with $\gtrsim 10^{6} M_{\Sol}$, 
the collapse of primordial clouds is induced 
by the H$_2$ cooling,
which sets a characteristic mass-scale of dense cores at 
several $100M_{\Sol}$ (Bromm, Coppi, \& Larson 1999; 
Abel, Bryan, \& Norman 2002). 
Inside these cores, protostars eventually form as 
a result of further graviational collapse by the H$_2$ cooling 
(Omukai \& Nishi 1998; Yoshida, Omukai, \& Hernquist 2008).
Although the exact value of final stellar mass 
is still elusive, it is speculated 
that the stars grow massive ($\gtrsim$ several $10M_{\Sol}$) 
owing to combined effects of high accretion rate and 
low opacity in primordial gas (Omukai \& Palla 2003; 
Bromm \& Loeb 2004; McKee \& Tan 2008; Hosokawa et al. 2011).
 
Despite with the same primordial composition, stars formed
under influences of radiative/mechanical 
feedbacks from pre-existing stars are expected 
to have a different characteristic mass-scale 
from that of the pop III.1 stars, and 
thus the terminology of pop III.2 stars have been 
proposed to specify them (O'Shea et al. 2008).   
Among various feedback mechanisms, e.g., 
photodissociation by the stellar far-ultraviolet radiation 
(Omukai 2001; O'Shea \& Norman 2008), the sweeping 
by supernova blast waves 
(Mackey, Bromm, \& Hernquist 2003; Salvaterra, Ferrara, \& Schneider 2004; Machida et al. 2005), 
one of the most well-studied is the case of star formation 
in relic HII regions of the first stars 
(O'Shea et al. 2005; Nagakura \& Omukai 2005; Machida, Omukai \& Matsumoto 2009).
The massive first stars ionize the surrounding material 
by UV radiation (e.g., Kitayama et al. 2004). 
Except two ranges ($10-40M_{\Sol}$ and $140-260M_{\Sol}$) 
in mass, they end their lives 
rather quietly without supernova explosions 
(Heger \& Woosley 2002; Umeda \& Nomoto 2002).
In those relic HII regions, the recombination proceeds in 
timescale $\sim 10^{8}$ yrs and the gas condenses again 
to form next-generation stars (Yoshida et al. 2007).  
Such pre-ionized primordial-gas follows a 
different thermal evolution from the pristine primordial gas 
of pop III.1 star formation 
(Uehara \& Inutsuka 2002; Nakamura \& Umemura 2002).
In the pre-ionized gas, 
more H$_2$ (with molecular fraction $\sim 10^{-2}$) than in the pristine gas 
($\sim 10^{-3}$)
forms via the electron-catalyzed reaction, 
\begin{eqnarray}
&{\rm H}& + {\rm e}  \rightarrow  {\rm H^{-} + \gamma }, \\
&{\rm H}& + {\rm H^{-}} \rightarrow  {\rm H_{2} + e }
\end{eqnarray}
and thus 
its cooling lowers the temerature below 
the minimum attainable in the pristine gas ($\simeq$ 200K).
In such a cold gas, HD fractionation is strongly 
favored via the exothermic (with energy difference
$\Delta E/k_{\rm B}=464$K) reaction
\begin{equation}
{\rm H_{2} + D^{+} \rightarrow HD + H^{+}}
\end{equation}
and the resultant HD cooling further lowers
the temperature to a few tens K, 
close to the CMB temperature in the high-redshift universe, 
$T_{\rm CMB}=27.3{\rm K}(1+z)/10$.
Those HD-cooling clouds fragment into dense cores of 
a few 10$M_{\Sol}$, which bear stars at their centers 
incorporating a large part of the material 
in the cores 
(Yoshida, Omukai \& Hernquist 2007; McGreer \& Bryan 2008).
Namely, the history of ionization in a 
low-density medium, or ``pre-ionization'', has a large effect 
on the subsequent thermal evolution of the metal-free gas, 
and possibly changes the characteristic stellar mass.  

For metal-enriched clouds, the pre-ionization effect 
has not been explored comprehensively so far.
Some previous studies, though, indicate that 
it has indeed effects on the 
thermal evolution at least in some range of metallicity.
For example, in the initially un-ionized cases with 
metallicity [Z/H] $\gtrsim -3.5$ 
\footnote{ [Z/H] $\equiv {\rm log}(Z/Z_{\Sol})$, 
where $Z$ is the mass fraction of metals in the medium.}, 
cooling by fine-structure transitions of carbon and oxygen 
lowers the temperature below the minimum value 
attainable solely by the H$_2$ cooling in the pristine gas 
($\simeq$ 200K at $\sim 10^{4}{\rm cm^{-3}}$)
(Omukai 2000; Bromm \& Loeb 2003).
On the other hand, from the pre-ionized initial condition, 
the minimum temperature attainable does not change 
so much from that in the primordial gas even for 
metallicity as much as [Z/H]$\sim -2$ (Jappsen et al. 2007). 

The pre-ionization effect in the metal-free gas
is largely due to the HD formation and cooling.
HD is also known to be quite susceptible 
to irradiation of FUV as well as ionization by the cosmic rays.
For example, even in the pre-ionized gas,
existence of only a weak FUV field prevents HD 
formation and its cooling (Yoshida, Omukai, \& Hernquist 2007b; 
Wolcott-Green \& Haiman 2010).
On the contrary, a certain amount of cosmic-ray (CR) 
irradiation (with a few \% of 
the energy density in the present-day Galactic disk) 
enables HD formation/cooling in an initially un-ionized cloud 
(Stacy \& Bromm 2007).

In this paper, we study whether the pre-ionization
affects the subsequent thermal evolution of low-metallicity 
clouds under a variety of intensities of 
the external FUV and CR fields.
We also aim to clarify the density range where the 
thermal evolution is influenced by the external 
FUV and CR irradiation.
The current prefered theory expects that 
transition from the Pop III 
massive star formation mode to the Pop II/I low-mass star 
mode is caused by the modification of thermal evolution 
due to the accumulation of metals in the star-forming gas,
albeit the exact value of the critical metallicity is still 
in dispute (Schneider et al. 2002; Bromm \& Loeb 2003; 
Dopcke et al. 2011).   
However, if the influences of environmental parameters, e.g., 
pre-ionization or external radiation fields, 
remain until high density where the fragmentation and low-mass
core formation occur, we also need to care about 
those parameters in 
studying fragmentation properties of low-metallicity clouds 
and thus the problem would become highly complex.

Our calculations here, fortunately, revealed that 
the pre-ionization alters the thermal evolution 
only in low density except cases with very 
low-metallicity ([Z/H] $\lesssim -4$) and 
under the low-FUV ($\lesssim 10^{-3}$)/CR fields ($\lesssim 0.1$ of the 
present-day Galactic disk levels). 
In addition, with a given metellicity, 
all the temperature evolution tracks 
with diffrent FUV/CR intensities converge 
before the density range where 
efficient cooling by dust triggers the cloud fragmentation. 
This demonstrates that in studying evolution of the dense cores, 
we need not care about the past ionization history except in 
the limited circumstances. 

This paper is organized as follows.
In Sec. 2, we describe the method of calculation and 
input physics in our model. In Sec. 3, the results for 
temperature evolution of low-metallicity clouds 
are presented and analysed. 
Finally, in Sec. 4, we conclude the paper 
by summarizing our findings.  

\section{Method of Calculation}
We calculate the thermal evolution of low-metallicity clouds 
by using a one-zone model, which is based on that by 
Omukai (2000) and Omukai et al. (2005), but with some modifications, 
including (i) chemical reaction network 
and its rate coefficients are updated 
and (ii) effects of FUV and CR irradiation, 
specifically, heating by 
photoelectric emission of dust grains and by CR ionization, 
and photo- and CR-induced chemical reactions, are added. 
The physical quantities, i.e., temperature, density, 
chemical fraction, etc., calculated in this model 
are regarded as those 
at the center of the cloud. 
The quantities calculated by this model are known to reproduce 
well the central evolution by 
hydrodynamical calculations for low-metallicity clouds 
(Omukai, Hosokawa, \& Yoshida 2010) as well as for 
primordial clouds under FUV irradiation
(Shang, Bryan, \& Haiman 2010). 

Without strong magnetic or turbulent supports, 
the dynamical collapse of self-gravitating clouds is 
well described by the Larson-Penston type similarity solution 
(Larson 1969; Penston 1968).
During the collapse, the cloud developes 
the core-envelope structure with homogeneous 
central region of nearly the Jeans length in size and 
the envelope where the density decreases with radius as 
$\propto r^{-2}$.
By mimicking the above property of the Larson-Penston solution, 
we assume the density increases by 
\begin{equation} 
\frac{d \rho}{dt}=\frac{\rho}{t_{\rm col}},  
\end{equation} 
where $\rho$ is the density of the gas, and the collapse 
timescale $t_{\rm col}=\sqrt{3 \pi/32 (1-f) G \rho}$ where $f$ is 
the ratio of pressure gradient to gravity and  
given as a function of the effective equation of state 
(see Omukai et al. 2005). 
The temperature evolution is followed by solving the 
energy equation:
\begin{equation} 
\frac{de}{dt}=-p \frac{d}{dt} \left(\frac{1}{\rho}\right)
- {\Lambda}_{\rm net},   
\label{eq:energy}  
\end{equation}
where the specific thermal energy is written as 
\begin{equation} 
e=\frac{1}{\gamma_{\rm ad} -1}\frac{k T}{ \mu m_{\rm H}}, 
\label{eq:defen} 
\end{equation} 
with the ratio of specific heat $\gamma_{\rm ad}$, 
the temperature $T$, and the mean molecular weight $\mu$,  
and, on the right-hand side, 
$p$ is the pressure
\begin{equation} 
p=\frac{\rho k T}{\mu m_{\rm H}},  
\end{equation}
and ${\Lambda}_{\rm net}$ is the net cooling rate per 
unit mass.
The cooling processes included in 
${\Lambda}_{\rm net}$ are the radiative cooling by 
the H Ly$\alpha$ transition, fine-structure line transitions of 
[CII], [CI], and [OI], and rotational 
(and also vibrational for H$_2$) 
transitions of molecules H$_{2}$, HD, CO, OH, and H$_2$O.
In calculating the H$_2$ cooling rate,
we also include the H$_2$-e and H$_2$-H$^+$ collisional 
transitions in addition to the H$_2$-H and 
H$_2$-H$_2$ collisions since, in a highly ionized gas, 
collisions with electrons and protons can potentially
dominate the H$_2$ excitation (Glover \& Abel 2008).
The newly added transition rates are taken from 
Draine, Roberge, \& Dalgarno (1983) for the H$_2$-e collisions 
and Gerlich (1990) for the H$_2$-H$^+$ collisions, following 
Glover \& Abel (2008). 
The cooling rates by the other species are 
calculated as in Omukai et al.(2010).
In evaluating the photon trapping effects,
we use the column density 
$N_{\rm H}=n_{\rm H} \lambda_{\rm J}$, 
recalling that the size of the core is about a Jeans length 
$\lambda_{\rm J}$. 
The heating processes are those associated with 
the H$_2$ formation, photoelectric emission from the dust grains,
and CR ionization.     
The heating rates by photoelectric emission of dust grains 
and by CR ionization are taken from Wolfire et al. (1995). 

The chemical reactions among species composed of 
elements H, He, C, and O are solved.
We consider the following 50 species as in Omukai et al.(2005): 
${\rm H}$, ${\rm H_2}$, ${\rm e^-}$, ${\rm H^+}$, ${\rm H_2^+}$, 
${\rm H_3^+}$, ${\rm H^-}$, ${\rm He}$, ${\rm He^+}$, ${\rm He^{++}}$, 
${\rm HeH^{+}}$, 
${\rm D}$, ${\rm D^{+}}$, ${\rm D^{-}}$, ${\rm HD}$, ${\rm HD^{+}}$, 
${\rm C}$, ${\rm C_2}$, ${\rm CH}$, ${\rm CH_2}$, 
${\rm CH_3}$, ${\rm CH_4}$, ${\rm C^+}$, ${\rm C_2^+}$, ${\rm CH^+}$, 
${\rm CH_2^+}$, ${\rm CH_3^+}$, ${\rm CH_4^+}$, ${\rm CH_5^+}$, ${\rm O}$, 
${\rm O_2}$, ${\rm OH}$, ${\rm CO}$, ${\rm H_2O}$, ${\rm HCO}$, 
${\rm O_2H}$, ${\rm CO_2}$, ${\rm H_2CO}$, ${\rm H_2O_2}$, ${\rm O^+}$, 
${\rm O_2^+}$, ${\rm OH^+}$, ${\rm CO^+}$, ${\rm H_2O^+}$, ${\rm HCO^+}$, 
${\rm O_2H^+}$, ${\rm H_3O^+}$, ${\rm H_2CO^+}$, ${\rm HCO_2^+}$ and
${\rm H_3CO^+}$.
The included reaction network among those species are updated 
from Omukai et al.(2005), whose reaction network is largely based 
on the UMIST database of Millar et al.(1997) in addition to the 
primordial-gas chemistry from several sources 
(e.g., Abel et al. 1997; Galli \& Palla 1998).
Our new network is based on the updated UMIST database of Woodall et al.(2007),
supplemented by primordial-gas chemistry of Glover \& Abel (2008).
Specifically, we construct our chemical network as follows.
First, we consider all the primordial-gas reactions in Glover \& Abel (2008) 
except those involving doubly deuterated species D$_2$ and D$_2^+$.
For uncertain three-body H$_2$ formation rate, 
we choose the value recommended by Glover (2008). 
Then, from a large number of the reactions listed in Woodall et al. 
(2007), we take those involving the above 50 species only, but excluding
the collider reactions, which become important only in higher density 
than our calculated range.
For overlapping reactions between the two compilations, 
we use the rate coefficient presented in Glover \& Abel (2008).
Primordial-gas reactions not in Glover \& Abel (2008) but in 
Woodall et al. (2007) are also considered.
Finally, photodissociation of H$_2$ and HD is added (see below).
Notable addition to our previous chemical model includes 
the photo- and cosmic-ray induced reactions as well as 
cosmic-ray induced photon reactions. 
The rate coefficients for CR-/photo-induced reactions by Woodall et al. (2007) 
are presented for those for the intensities in the Galactic disk, 
i.e., the CR primary ionization rate  
$\zeta_{\rm disk}=1.3 \times 10^{-17}{\rm s^{-1}}$, and 
the FUV intensity with respect to the Draine field 
$G_{0}=1.7$ (Draine 1978), where 
the Habing parameter $G_{0}$ is defined by the FUV 
flux in the range 6-13.6 eV normalized by 
$1.6 \times 10^{-3} {\rm erg~cm^{-2}~s^{-1}}$ (Habing 1968).
We thus rescale the coefficients in Woodall et al. (2007) 
for CR-induced reactions 
and photo-ionization/dissociation by multiplying the factor 
$\zeta/\zeta_{\rm disk}$ and $G_{0}/1.7$, respectively. 
This scaling corresponds to the assumption that the FUV spectrum is 
the same as the Galactic field, which can be invalidated, for example, 
for the radiation field dominated by massive pop III stars with 
high ($\sim 10^5$ K) effective temperature. 
The photo-chemical reaction coefficients decrease
exponentially with the visual extinction $A_{\rm V}$
owing to the shielding by dust grains.
The visual extinction is related with the column density 
of the hydrogen nuclei as 
$A_{\rm V}=5.3 \times 10^{-22} N_{\rm H}{\rm cm^{-2}} 
(Z/Z_{\Sol})$ under the assumption of grains with 
the same optical properties as in the solar neighborhood 
(Bohline et al. 1978; Rachford et al. 2009).
For the photodissociation of H$_{2}$ and HD, and CO, 
in addition to the dust shielding, the shielding by H$_2$ becomes important.
For H$_2$ photodissociation, we use the reaction rate by 
Draine \& Bertoldi (1996) 
\begin{equation}
k_{\rm H_2 pd}=4.50\times 10^{-11}f_{\rm sh} {\rm exp}(-2.5 A_{\rm V}) G_{0}~{\rm s^{-1}},
\label{eq:H2pd}
\end{equation}
where the shielding factor $f_{\rm sh}$ are taken from 
Wolcott-Green \& Haiman (2010).
For the HD photodissociation, we use the same expression as H$_2$ 
(eq. \ref{eq:H2pd}), but the proper shielding factor for HD 
(Wolcott-Green \& Haiman 2010).
For the photodissociation of CO, 
we modify the reaction coefficients from Woodall et al. (2007) 
by including the shielding by H$_{2}$.
We use the fit to Lee et al. (1996)'s results 
by Hosokawa \& Inutsuka (2006), which is within 20 \% accuracy for 
$N_{\rm H_2} < 8 \times 10^{21}{\rm cm^{-2}}$ although substantially worse 
at higher $N_{\rm H_2}$:
\begin{equation}
k_{\rm CO pd}=2.0\times 10^{-10}
{\rm exp}\left(-3.5A_{\rm V}-\frac{N_{\rm H_2}}{1.6 \times 10^{21}{\rm cm^{-2}}}\right) (G_{0}/1.7)~{\rm s^{-1}}.
\end{equation} 
We do not take the shielding of CRs into account. 
Although the CR shielding becomes important for column density 
$\gtrsim 29 {\rm g~cm^{-2}}$ (Umebayashi \& Nakano 1986), 
the column density does not reach such high 
and the shielding does not have a large effect in our 
calculations.

The initial number density is taken at $0.1{\rm cm^{-3}}$, 
which is a typical value in the relic HII regions of the first 
stars (Yoshida et al. 2007).
We consider two types of the initial conditions: 
(i) un-ionized condition: the initial temperature 
$T_{\rm ini}=1000{\rm K}$, the ionization degree 
$y_{\rm ini}({\rm e})=3 \times 10^{-4}$, 
molecular fraction $y_{\rm ini}({\rm H_2})=2 \times 10^{-5}$, 
and (ii) pre-ionized condition : $T_{\rm ini}=8000{\rm K}$, 
$y_{\rm ini}({\rm e})=0.5$, $y_{\rm ini}({\rm H_2})=0$. 
The un-ionized condition (i) is typical for the formation of the first stars (Yoshida et al. 2006), while the pre-ionized condition (ii) corresponds to the transitional state where the recombination is proceeding in the relic HII region (Yoshida et al. 2007a).
The factor of a few variations in those initial chemical 
abundances do not alter the thermal evolution of gas 
presented below. 

\section{Results}
\subsection{No FUV/CR Cases}
First, we see the cases with neither CR nor FUV field. 
Figure 1 presents the comparison of thermal evolution 
of low-metallicity gas in the un-ionized and 
pre-ionized cases.

The lower panel shows the results in un-ionized cases. 
In low densities ($\lesssim 10^{4}{\rm cm^{-3}}$), 
temperatures are almost the same for the clouds with [Z/H]$ \leq -4$ 
as H$_2$ is the dominant coolant in this density/metallicity range.  
For higher metallcity, i.e., [Z/H]$\geq -3$, 
cooling by the C and O fine-structure transitions breaks 
this degeneracy: the temperature becomes lower 
for higher metellicity.
In the higher density range, temperature tracks deviate 
from the $Z=0$ one also for [Z/H] $\lesssim -4$, 
owing to cooling by H$_2$ formed by the dust surface reaction, 
and by H$_2$O, both of whose amounts are dependent on metallicity. 

Next we see the pre-ionized cases (upper panel).
For [Z/H]$\lesssim -4$, HD dominates the cooling in $\lesssim 150$K, 
and thus the temperature tracks degenerate until 
$\sim 10^{8}{\rm cm^{-3}}$ where the gas is heated up 
to $\sim 300$K by the exothermic three-body H$_2$ forming 
reaction.
In comparison with the un-ionized clouds, which 
cool solely by H$_2$, the temperature in the pre-ionized clouds 
is significantly lower in the metallicity range [Z/H]$\lesssim -4$,
thanks to the HD cooling.
This effect, efficient HD cooling in the pre-ionized gas,  
is well-known for the primordial gas, i.e., 
the pop III.2 star formation.  
For $\gtrsim 10^{8}{\rm cm^{-3}}$, the warm environment makes 
H$_2$ the more important coolant than HD, and thus 
the temperature tracks become the same as those 
in the un-ionized cases.  
With more metals ([Z/H] $\geq -3$),
the cooling rate by the fine-structure lines exceeds 
that by HD. The temperatures are then similar in both 
un-ionized and pre-ionized cases, although 
slightly lower in the pre-ionized cases. 
In the case of [Z/H]$=-1$, the pre-ionized gas takes slightly longer time, i.e., higher density, to reach the CMB temperature of 30K. This just reflects 
the higher initial temperature in the pre-ionized case.  

To summarize, below the metallicity [Z/H]$\simeq -3$, 
HD is the dominant coolant for $\lesssim 10^{8} {\rm cm^{-3}}$ 
in the pre-ionized gas, and the thermal evolution 
depends on the initial ionization state as is known 
for the pop III star formation. 
For higher metallicity, on the other hand, 
the fine-structure-line cooling becomes more important 
and the pre-ionization has little impact on the subsequent 
thermal evolution.  

\begin{figure*}
  \begin{center}
    \FigureFile(150mm,150mm){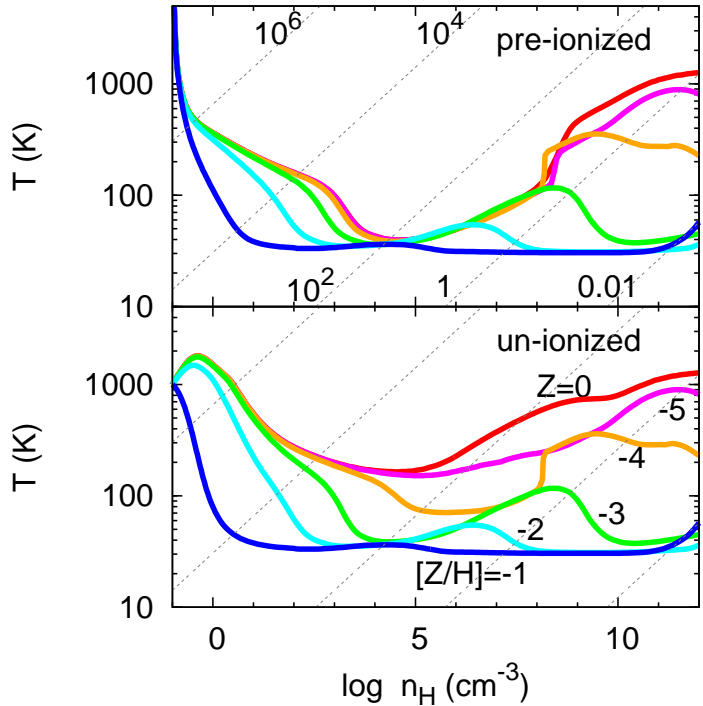}
  \end{center}	
\caption{
Temperature evolution in clouds with six different metallicities 
($Z$=0 ({\it red}), [Z/H]=-5 ({\it magenta}), -4 ({\it orange}), -3 ({\it green}), -2 ({\it cyan}), and -1 ({\it blue}))
against number density 
in the pre-ionized (upper panel) and for un-ionized (lower panel) cases. 
The cases with neither FUV nor CR incidation. 
The values of metallicities [Z/H] are indicated by numbers in the lower panel. 
The thin dashed lines indicate the constant Jeans mass, 
whose value is shown by numbers in the upper panel.
\label{fig:fig1}}
\end{figure*}

\subsection{Cases only with FUV irradiation}
\begin{figure*}
  \begin{center}
    \FigureFile(200mm,200mm){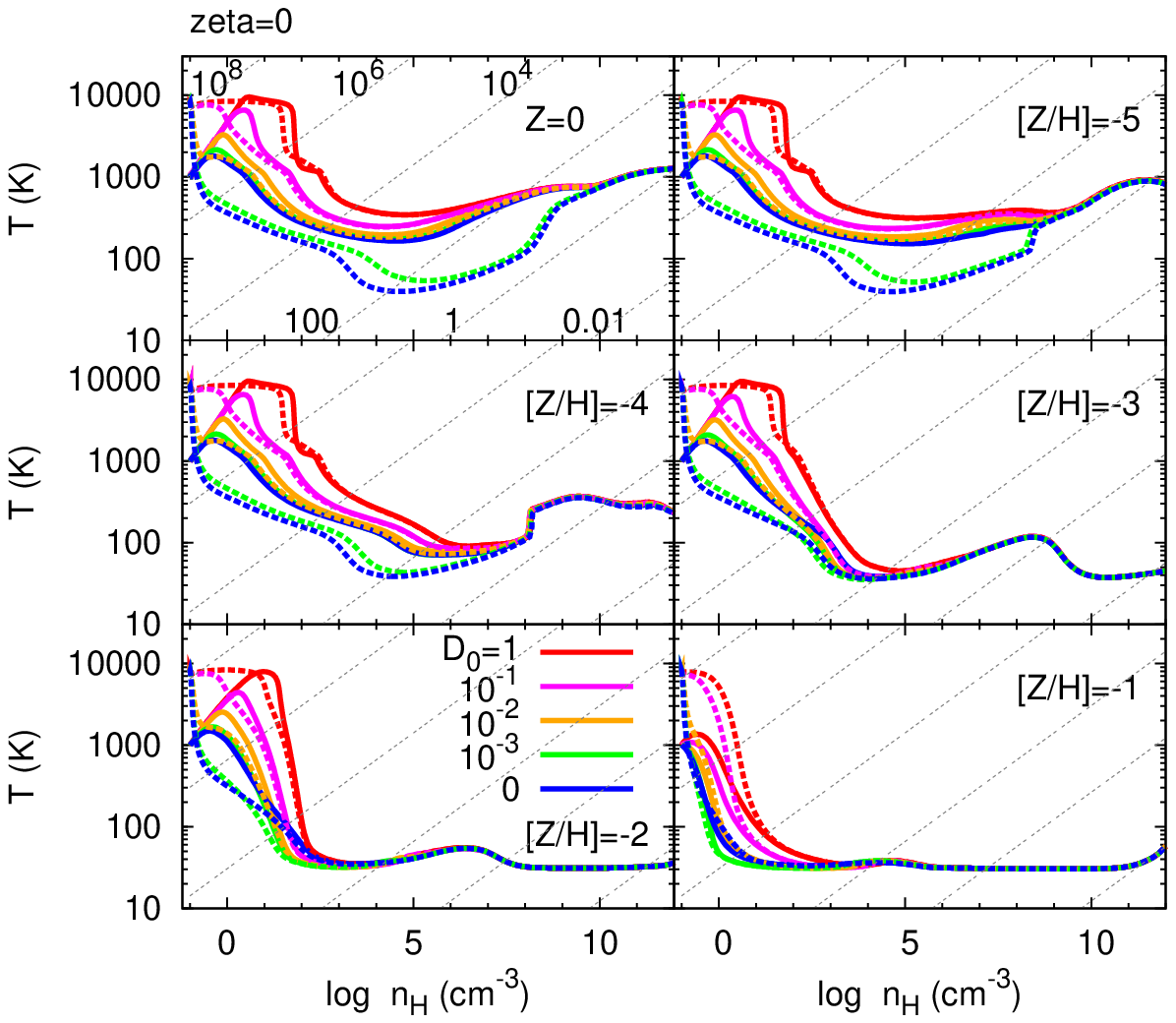}
  \end{center}	
\caption{Thermal evolution of low-metallicity clouds 
with FUV irradiation, but without CR incidation.
Panels correspond to the cases with 
different metallicities $Z=0$ ({\it left-top}), 
${\rm [Z/H]}=-5$ ({\it right-top}), $-4$ ({\it left-middle}), 
$-3$ ({\it right-middle}), $-2$ ({\it left-bottom}) and 
$-1$ ({\it right-bottom}).
In each panel, the solid and dashed lines indicate 
the un-ionized and pre-ionized cases, respectively, 
for five different FUV strengths, 
$D_{0}\equiv G_{0}/1.7=0$ ({\it blue}), $10^{-3}$ ({\it green}), 
$10^{-2}$ ({\it orange}), $10^{-1}$ ({\it magenta}) and  
1 ({\it red}). 
The thin dashed lines indicate the constant Jeans mass, 
whose value is shown by numbers in the upper-left panel.
\label{fig:fig2}}
\end{figure*}

Next we see the cases with external FUV irradiation
but without CR incidation. 
Panels in the Figure \ref{fig:fig2} correspond to the cases with 
different metallicities $Z=0$ ({\it left-top}), 
${\rm [Z/H]}=-5$ ({\it right-top}), $-4$ ({\it left-middle}), 
$-3$ ({\it right-middle}), $-2$ ({\it left-bottom}) and 
$-1$ ({\it right-bottom}).
In each panel, the solid and dashed lines indicate 
the un-ionized and pre-ionized cases, respectively, 
for five different FUV strengths, 
$D_{0}\equiv G_{0}/1.7=0$ ({\it blue}), $10^{-3}$ ({\it green}), 
$10^{-2}$ ({\it orange}), $10^{-1}$ ({\it magenta}) and  
1 ({\it red}).
In the pre-ionized cases ({\it dashed}) with strong 
FUV fields ($D_{0}\geq 0.1$), 
heating by the photoelectric emission and 
cooling by the Ly$\alpha$ emission balances, 
and the temperature remains 
almost constant at the initial value of 8000K for a while. 
In the un-ionized cases ({\it solid}), the temperature 
increases adiabatically from the initial 1000K 
to several thousand K, depending on the strength of FUV field, 
since the photodissociation prevents the H$_2$ formation and its cooling.  
In both un-/pre-ionized cases, the temperature begins to fall down
when the H$_2$ self-shielding against the photodissociation 
becomes effective for [Z/H]$\lesssim-3$.
With high enough metallicity [Z/H]$\gtrsim-2$, 
the fine-structure-line cooling causes the temperature decrease 
before the H$_2$ is self-shielded.  

As discussed in Sec. 3.1, without an FUV field, 
the thermal evolution of low-metallicity gas ([Z/H] $\lesssim -4$) 
bifurcates according to its initial ionization state. 
This still holds true in weak enough FUV cases (see the cases with 
$D_{0} \leq 10^{-3}$).
With higher FUV intensity ($D_{0} \geq 10^{-2}$), 
however, the temperatures 
in un-ionized and pre-ionized cases quickly converge each other 
during the initial temperature-decreasing phase.
The dependence on the initial ionization state immediately 
disappears.
This is because sufficient HD for cooling is not formed in 
either un-ionized nor pre-ionized case:
even in the pre-ionized cases, H$_2$ photodissociation 
by FUV field prevents the gas to cool $\lesssim$ 150K, thereby quenching 
HD formation, which proceeds only in the low-temperature environments.
For [Z/H] $\gtrsim -3$, as in no UV cases, 
the minimum temperatures are the same for pre-/ un-ionized cases 
and no effect of pre-ionization is observed during the 
subsequent thermal evolution.

It should be noted that, in all the cases, 
for $\gtrsim 10^8{\rm cm^{-3}}$, 
neither the initial ionization state nor the 
irradiating FUV strength affect the thermal evolution. 
This is because the H$_2$ fraction boosts up at 
$\sim 10^8{\rm cm^{-3}}$ by the three-body reaction 
and its differences due to different initial 
ionization or FUV field strength in lower densities 
are wiped out.
For high metallicity cases [Z/H] $\geq -3$, 
this convergence of the thermal evolution 
occurs at even lower density since the H$_2$ cooling is only 
subordinate to the fine-strucutre-line cooling.

\subsection {Cases also with CRs}
Finally, we see the cases not only with FUV irradiation, but
also with the CR incidation.
In the solar neighborhood, the CR energy density is estimated, 
in terms of the primary ionization rate, to be  
$\zeta_{\rm disk} \sim 10^{-17} {\rm s^{-1}}$ 
(Webber \& Yushak 1983), 
with about an order of magnitude uncertainty 
(e.g., Indrilo et al. 2007).
Although the CR strength is, of course, far more
uncertain in the early universe, about 1/100 of 
the Galactic value  
$\zeta \sim 10^{-19}{\rm s^{-1}}$ is claimed as an 
intergalactic CR field (Stacy \& Bromm 2007).
In this section, we regard the CR intensity as a free paremeter 
and study cases with $\zeta = 10^{-19}...10^{-17} {\rm s^{-1}}$. 

\begin{figure*}
  \begin{center}
    \FigureFile(200mm,200mm){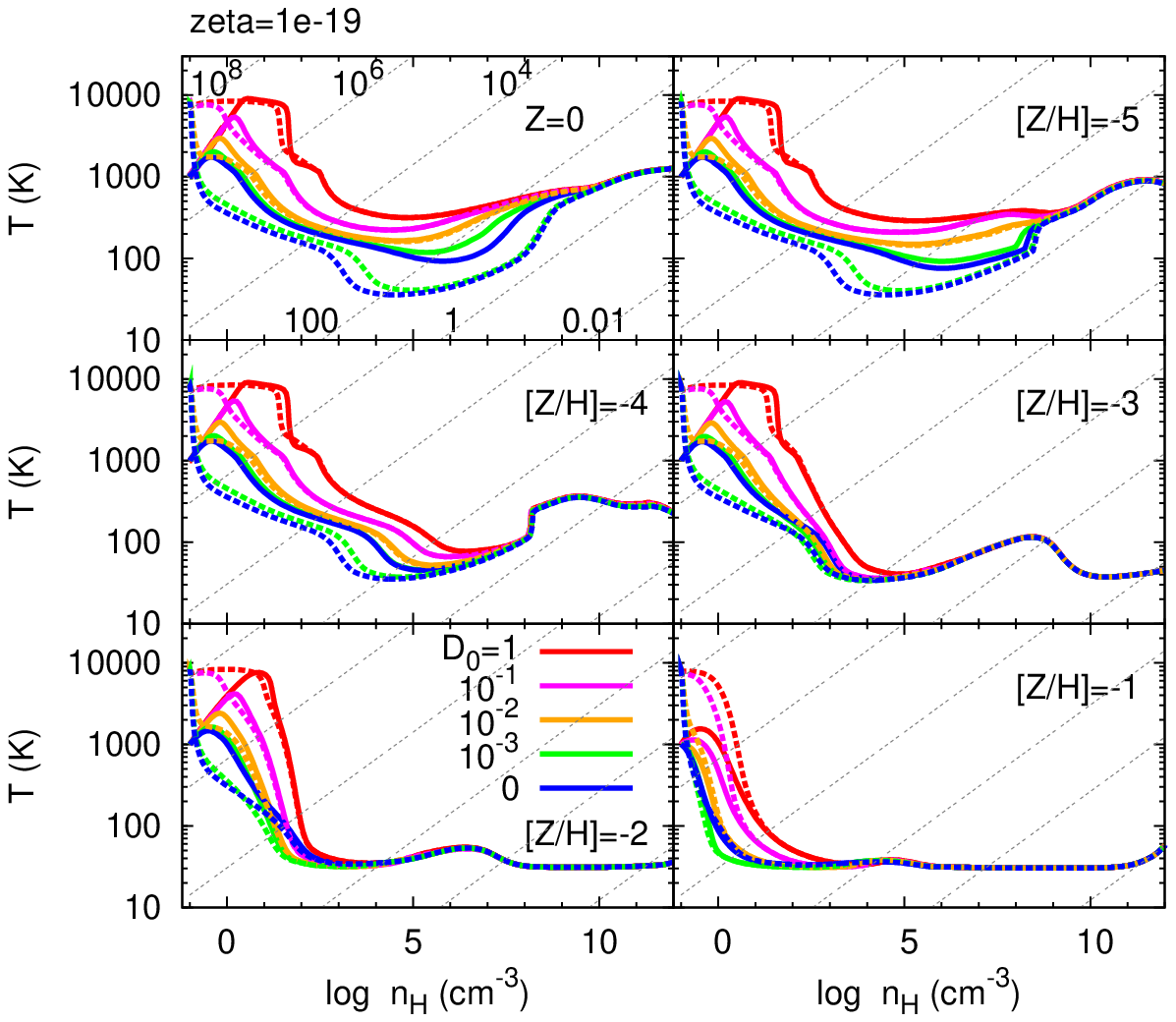}
  \end{center}	
\caption{Same as Figure \ref{fig:fig2} 
but with CR indication at the cosmic-ray primary ionization rate 
$\zeta=10^{-19}{\rm s^{-1}}$.
\label{fig:fig3}}
\end{figure*}

\begin{figure*}
  \begin{center}
    \FigureFile(200mm,200mm){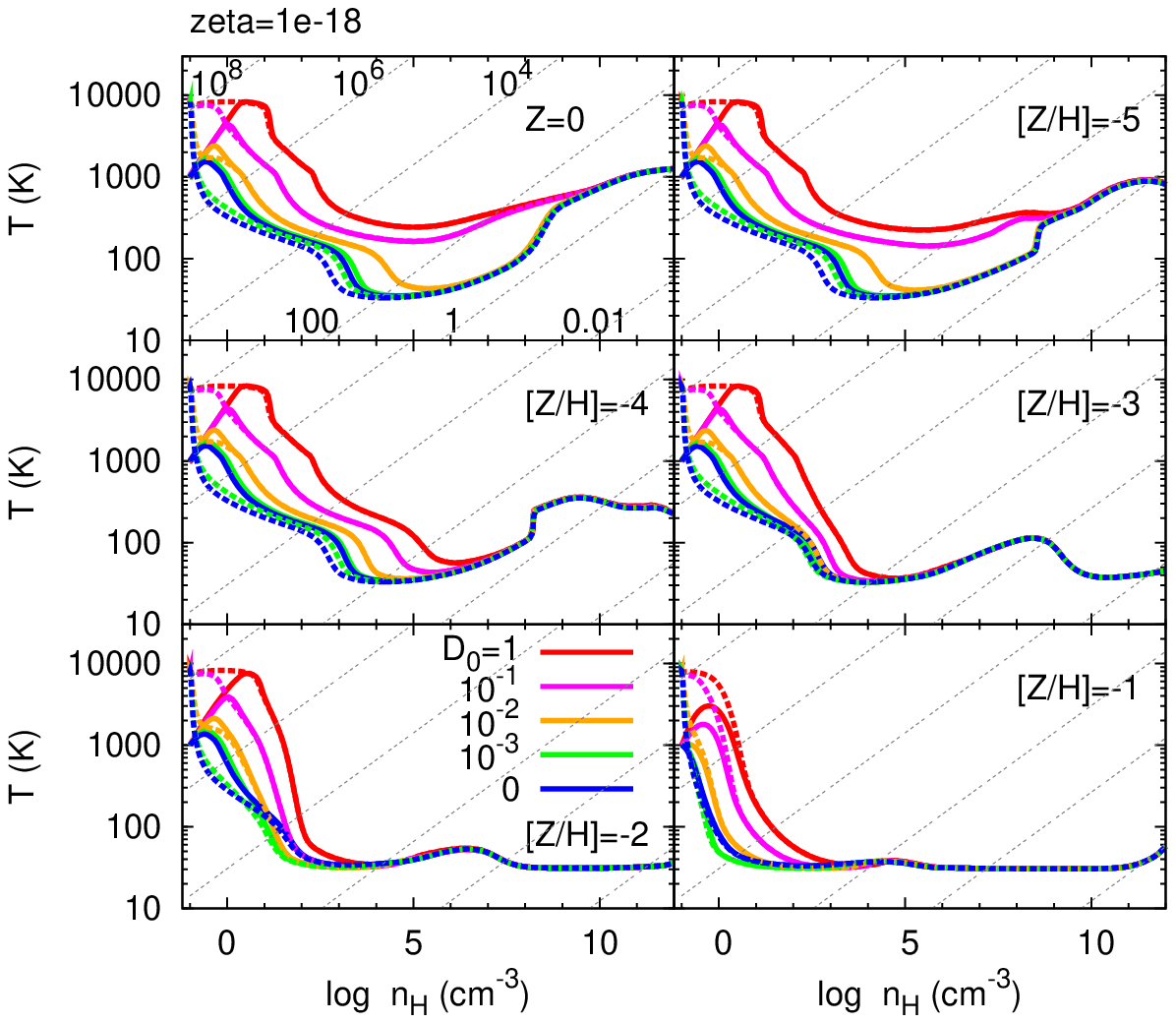}
  \end{center}	
\caption{Same as Figure \ref{fig:fig2} but for 
$\zeta=10^{-18}{\rm s^{-1}}$.
\label{fig:fig4}}
\end{figure*}

\begin{figure*}
  \begin{center}
    \FigureFile(200mm,200mm){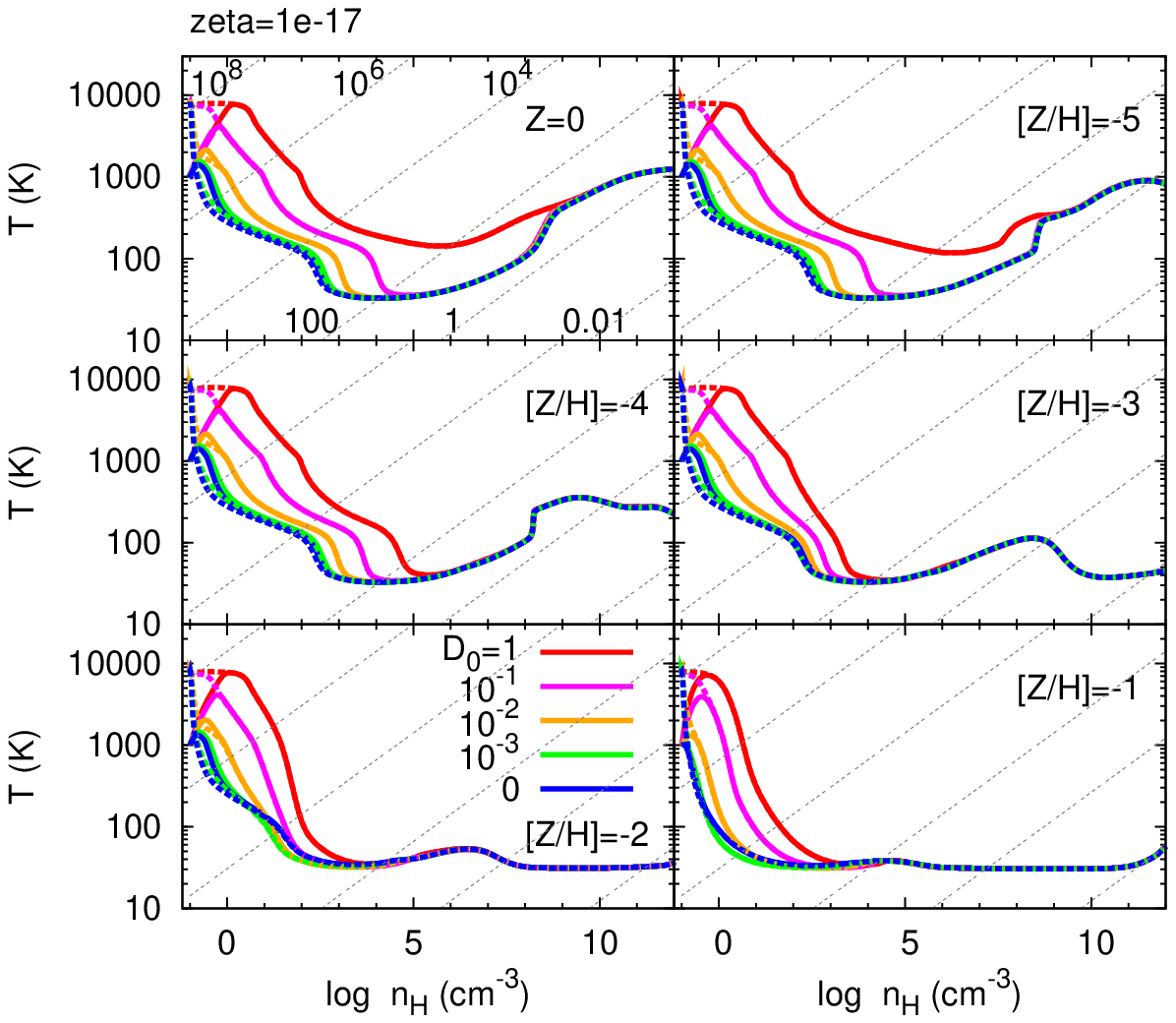}
  \end{center}
\caption{Same as Figure \ref{fig:fig2} but for 
$\zeta=10^{-17}{\rm s^{-1}}$.
\label{fig:fig5}}
\end{figure*}

Figure \ref{fig:fig3} shows the weak CR cases of 
$\zeta= 10^{-19} {\rm s^{-1}}$. 
At first glance, these results resemble very well with 
those without CRs shown in Figure \ref{fig:fig2}.
However, at very low-metallicity [Z/H] $\lesssim -4$ 
and at the same time 
with low FUV intensity $D_{0}\lesssim 10^{-3}$, 
the temperature in the un-ionized cases
is somewhat lower in the range of 
$10^{5}-10^{7} {\rm cm^{-3}}$ than in the no CR cases, 
because the CR ionization enables the HD cooling 
(c.f. Stacy \& Bromm 2007).
The cases with higher CR intensity 
$\zeta= 10^{-18}$ and $10^{-17} {\rm s^{-1}}$ are presented 
in Figures \ref{fig:fig4} and \ref{fig:fig5}, respectively. 
The latter CR value is as high as in the present-day 
Galactic disk. 
Most notably, with increasing CR strength, 
the pre-ionization effect becomes weaker. 
Without CR, 
the pre-ionization effect is clearly visible 
for clouds with low-metallicity [Z/H] $\lesssim -4$ and 
simultaneously with low FUV intensity $D_{0}\lesssim 10^{-3}$ 
(Sec. 3.2). 
With increasing CR intensity, the differences 
among the pre-ionized and un-ionized cases 
diminish even in those cases.  
Recall that, without CR, the HD formation/cooling is operative
only in the pre-ionized environment.  
With some CR incidation, CR ionization induces the 
HD formation/cooling even in the un-ionzed condition. 
The differences in temperature 
evolution among pre-ionized and un-ionized cases thus 
become smaller for higher CR strength.  
Besides ionization, CRs also have the heating effect. 
This is clearly seen in Figures  \ref{fig:fig4} and 
\ref{fig:fig5} as a rapid initial temperature 
rise in the un-ionzed cases with metallicity [Z/H]$\lesssim -2$.  
For density $\gtrsim 10{\rm cm^{-3}}$, this effect is not visible 
compensated by efficient cooling. 

\section{Summary and Conclusion}
We have studied the effect of pre-ionization, 
as well as the far-ultraviolet irradiation and cosmic-ray 
incidation, on the thermal evolution of low-metallicity 
star-forming gas. 
Our findings can be summarized as follows:

1. Without external FUV or CR irradiation, 
HD formation is promoted in the initially ionized 
(``pre-ionized'') clouds. 
This HD cooling makes the temperature in low-metallicity 
([Z/H] $\lesssim -3$) 
clouds fall below the minimum value in 
the un-ionized gas attainable solely by H$_2$ cooling ($\simeq$ 200K).
The temperature finally reaches a few 10 K, similar
to the high-redshift CMB temperature. 
With more metals, regardless of HD formation/cooling, 
the gas is able to reach such low temperature
by the fine-structure line cooling.
In this case, the pre-ionization does not have impact 
on the thermal evolution. 

2. Presence of a weak FUV field ($G_{0} \sim 10^{-2})$ 
prevents the HD formation/cooling even in the pre-ionized gas. 
This is because of the H$_2$ photodissociation:
without sufficient H$_2$, the gas does not cool 
to $\lesssim 150$K, where the deuterium is rapidly converted to HD.
In this case, the pre-ionization does not affect 
the subsequent thermal evolution of low-metallicity gas. 

3. The cosmic-ray ionization as high as 
$\zeta \sim 10^{-18}{\rm s^{-1}}$ in terms of the primary 
ionization rate, or 
$\sim 10\%$ of the solar neiborhood level, 
enables HD formation/cooling even 
in the un-ionized gas, thereby eliminating
the pre-ionization effects even in very 
low-metallicity cases.

In summary, the pre-ionization has remarlable effects only 
in clouds with very low metallicity ([Z/H] $\lesssim -3$), 
and under low FUV ($G_{0} \lesssim 10^{-2}$) or
low CR ($\zeta \lesssim 10^{-18}$) fields. 

In addition, 

4. the effects of external FUV/CRs on thermal evolution 
vanishes until $10^{8}{\rm cm^{-3}}$ in all the cases we studied.  

The convergence of thermal evolution at high density has 
important consequence on the mass-scale of forming stars.
In the framework of dynamical fragmentation, clouds fragments 
during efficient cooling phase, where 
the temperature decreases with density, while fragmentation 
hardly occurs in the temperature-increasing phase 
(Larson 1985, 2005; Li et al. 2003).
The characteristic fragmentation mass-scales are thus set by 
the Jeans mass at the local minima of temperature. 
The low-metallicity clouds with $-5 \lesssim {\rm [Z/H]} \lesssim-1$ has 
two temperature minima during the collapse; 
the first one in the lower-density 
($\sim 10^{3}-10^{5} {\rm cm^{-3}}$)
due to the metal or H$_2$, HD line cooling, and the second one 
in the higher-density ($\sim 10^{10}-10^{15} {\rm cm^{-3}}$) 
due to the cooling by dust thermal emission. 
Since the location of the first lower-density temperature minimum 
is vulnerable to the external FUV radiation, 
the fragmentation mass-scale owing to the line cooling 
can vary from cloud to cloud depending on the intensity of 
surrounding FUV field.
On the other hand, as the temperature evolution depends 
neither on the pre-ionization nor external FUV radiation 
for $\gtrsim 10^{8} {\rm cm^{-3}}$, 
the location of the temperature minimum by the dust cooling 
is robust.
Thus, in studying the fragmentation by dust cooling 
we do not need to care about the external radiation.
The results of numerical simulations for dust-induced 
fragmentation (e.g., Tsuribe \& Omukai 2006, 2008; 
Clark et al. 2008; Dopcke et al. 2011)
remain valid also in the external radiation field.  


The author is very grateful for constructive comments by anonymous 
referee.
This study is supported in part by the
Grants-in-Aid by the Ministry of Education, Science and Culture of
Japan (2168407, 21244021). 


\end{document}